\documentstyle[12pt]{article}
\def\be{\begin{equation}}
\def\ee{\end{equation}}
\def\real{{\rm I\!R}}

\begin{document}
\begin{titlepage}

\title{Hand-waving Refined Algebraic Quantization
    \thanks{
            Talk given at the meeting {\it Modern Modified Theories
            of Gravitation and Cosmology}, 
            Ben Gurion University, Beer Sheva, June 29--30, 1997. 
                     }
                          }

\author{Franz Embacher\\
        Institut f\"ur Theoretische Physik\\
        Universit\"at Wien\\
        Boltzmanngasse 5\\
        A-1090 Wien\\
        \\
        E-mail: fe@pap.univie.ac.at\\
        \\
        UWThPh-1997-22\\
        gr-qc/9708016 
                      } 
\date{}

\maketitle

\begin{abstract}
Some basic ideas of the Refined Algebraic Quantization scheme 
are outlined at an intuitive level, using a class
of simple models with a single wave equation as quantum constraint. 
In addition, hints are given how the scheme is applied to 
more sophisticated models, and it is tried to make transparent 
the general pattern characterizing this method. 
\end{abstract}

\end{titlepage}

\section{Introduction}

In a slight misuse of the purpose of this meeting, I will not 
talk about a ``modified'' or ``alternative'' theory of gravity 
but rather give a brief introduction into a 
``modified'' or ``alternative'' 
quantization method. 
Although currently being applied with considerable success 
to the program of quantizing full gravity, 
this method seems to be unknown to a large fraction of the 
relativity community. 
Hence, although I am not a very specialist on this topic, 
let me please take the opportunity to present the 
basic ideas of 
what is called Refined Algebraic Quantization (RAQ). 
I will put emphasis on some of its 
aspects that can be understood intuitively 
and present them in a hand-waving fashion, 
rather than being mathematically very rigorous.  
In case I succeed in stimulating some interest, 
the existing literature will certainly be helpful in 
getting further. 
\medskip

The RAQ scheme deals with constrained systems. 
The type of constraints most challenging for gravitational theorists and 
cosmologists is of course the one appearing in the Hamiltonian 
formulation of general relativity: quadratic in the momenta. 
The recent development of the RAQ scheme is to some extent linked with 
the program of loop quantum gravity\cite{Aetal}  
(the basic idea dating back
to the Sixties 
\cite{Nachtmann} and having been re-invented 
in the Nineties 
\cite{Higuchi,Landsmanetal,Marolf1}). I will ignore the 
``loop'' issue 
(and also the issue of the $SU(2)$ connection and the famous
spin networks appearing there) 
since, logically, quantum gravity just provides an area of 
application of the general set of rules called RAQ. Rather than 
stating these rules precisely in their most sophisticated form, 
I will first develop an easy version by treating a class of 
drastly simplified models. In these models almost all features
of the ``true'' gravity constraints have disappeared, except for 
one: being quadratic in the momenta. Only afterwards, I will 
provide some hints towards the exactification of the scheme and 
its generalization to more realistic systems. 
\medskip 

In other words, I consider what may be called minisuperspace 
models or, mathematically equivalent, the motion of a 
scalar particle in a space-time background. 
In such a scenario, one has a fixed background 
structure, consisting of a finite dimensional manifold $\cal M$, 
a Pseudo-Riemannian metric 
$ds^2 = g_{\alpha\beta}\,d x^\alpha d x^\beta$ 
(signature $-+\dots+$) and a real function $U$. 
Choosing a coordinate system $(x^\alpha)$ in $\cal M$, 
the classical phase space locally consists of pairs 
$(x^\alpha,p_\alpha)$. There is only one classcial 
constraint, reading
\be 
{\cal C} \equiv p_\alpha p^\alpha + U = 0\,. 
\label{C}
\ee
Upon substituting $p_\alpha\rightarrow - i \partial_\alpha$ 
and choosing the simplest covariant operator ordering, 
the classical constraint is promoted into the 
quantum constraint (wave equation) 
\be
\widehat{\cal C }\,\psi \equiv 
(- \nabla_\alpha \nabla^\alpha + U ) \psi = 0 \, , 
\label{WDW}
\ee
the solutions of which we call physical states. 
(This will be slightly modified and made more precise later on). 
\medskip 

In a quantum cosmological context, ${\cal M}$ is the minisuperspace
manifold 
and equation (\ref{WDW}) is the corresponding Wheeler-DeWitt equation. 
If $\cal M$ denotes a space-time manifold, equation (\ref{WDW}) is the 
Klein-Gordon equation for a scalar particle moving in this 
space-time and feeling a potential $U$. 
Hence, one and the same mathematical equation (\ref{WDW}) may be 
interpretated in two ways that are entirely different from 
each other from the point 
of view of physics. (In particular, the notion of a class of 
observers moving on a congruence of world-lines in space-time 
and being freely adjustable 
has no counterpart in minisuperspace). 
I stress this point, because the 
RAQ scheme may look very strange at first sight when applied to 
a particle quantization scenario, and it may easier loose this 
alien appeal 
when a quantum gravitational (or simpler: a quantum cosmological) 
context is kept in mind. 
\medskip

So let us start with the wave equation (\ref{WDW}). 
In the traditional 
Klein-Gordon quantization one is guided by the idea of $\psi$ being 
the wave function of a particle. The key object to give a 
suitably defined space of solutions of (\ref{WDW}) some 
additional structure is the indefinite scalar product 
\be
Q(\psi_1,\psi_2)= -\,\frac{i}{2}\, \int_\Sigma d\Sigma^\alpha\,
(\psi_1^*\stackrel{\leftrightarrow}{\nabla}_\alpha \psi_2) \, , 
\label{KG}
\ee
where $\Sigma$ is a spacelike hypersurface 
(with sufficiently regular asymptotic behaviour). 
The integrand of $Q$ is a conserved current, upon which 
many of the physical interpretations of $\psi$ are based. 
\medskip 

Let us recall that for the case of a massive particle 
in Minkowski space ($g_{\alpha\beta}=\eta_{\alpha\beta}$ being flat 
and $U=m^2$ being constant) there is a 
unique decomposition of any (asymptotically well-behaved) wave function 
into a positive and a negative frequency part
\be
\psi(t,\vec{x}) = \int 
\frac{d^3k}{ (2\pi)^{3/2} \sqrt{  \omega({\vec{k}})} }
\Big( 
a(\vec{k}) e^{i (\vec{k}\vec{x}- \omega(\vec{k}) t)} + 
b(\vec{k}) e^{i (\vec{k}\vec{x}+ \omega(\vec{k}) t)} \Big) 
\equiv \psi_+(t,\vec{x})  + \psi_-(t,\vec{x}) \, ,  
\label{posneg} 
\ee
where $\omega(\vec{k})=\sqrt{\vec{k}^2+m^2}$. 
The scalar product (\ref{KG}) then reads 
\be
Q(\psi_1,\psi_2) = 
\int d^3k \Big( a^*_{1}(\vec{k}) a_2(\vec{k}) - 
b^*_{1}(\vec{k}) b_2(\vec{k})  \Big) \, ,  
\label{KGmink}
\ee
the minus sign nicely illustrating the indefiniteness of $Q$ 
and the way how actually {\it two} Hilbert space structures 
(one with positive and one with negative sign) 
emerge from (\ref{WDW}). In the early approaches to 
the quantum theory of relativistic particles, they would have been 
called one-particle (and one-antiparticle) Hilbert space, 
respectively. 
\medskip  

On a curved background, an analogous construction becomes ambigous, 
unless there is a local symmetry with timelike trajectories, thus 
providing a preferred time coordinate with respect to which 
the frequency decomposition is defined. (In the 
Minkowski space example above, one of course assumes the time 
coordinate of any intertial frame to do this job.
Due to Lorentz invariance,  
the resulting decomposition (\ref{posneg}) 
is independent of the frame chosen). 
This ambiguity in quantizing a particle in a generic curved 
background just means that the definition of what is a particle 
and what is an antiparticle depends on the observer(s). 
This is not just a shortcome of the quantization method 
but reflects physical properties: it is 
the origin of particle production in curved space-time.  
However, in lack of a preferred expansion like 
(\ref{KGmink}) --- which displayed two Hilbert space
structures --- it is not at all easy to say which of the 
functions $\psi$ on $\cal M$ solving the wave equation shall 
qualify as physical states and which shall not, and 
usually one has to worry about cumbersome boundary conditions.
\medskip 

The modern way to carry out the particle quantization program 
in a curved background is not this approach but second 
quantization. However, quantum cosmology introduces 
a different physical interpretation of the ingredients of 
(\ref{WDW}), so that one must reconsider the quantization method. 
Here, the RAQ scheme may be brought into the game as a method 
which, in minisuperspace models of the type (\ref{WDW}), 
generates a well-defined Hilbert space of physical states 
in a much more direct way than the Klein-Gordon approach. 
Also in realistic models, containing more than 
just one constraint, this scheme seems to be more likely to work 
than one based on a generalization of (\ref{KG}). 
\medskip 

\section{Delta function acrobatics} 

In contrast to the Klein-Gordon quantization based on 
(\ref{KG}), the RAQ relies on the positive
definite inner product 
\be
\langle\psi_1,\psi_2\rangle 
=\int_{\cal M} d^n x\,\sqrt{-g}\,\psi_1^*\psi_2 
\label{KIN}
\ee 
on the Hilbert space ${\cal H}$ of square-integrable functions 
on $\cal M$. Although this integral will in general not converge if 
solutions of (\ref{WDW}) are inserted, the main motivation 
driving one away from the traditional scheme is to reject any indefinite 
scalar product and to retain (\ref{KIN}) as the starting point. 
\medskip 

Proceeding heuristically, I write down a solution 
to the quantum constraint (\ref{WDW}) in a very formal way. 
Let us insert the constraint operator $\widehat{\cal C}$ 
from (\ref{WDW}) 
into the delta function and arrive at the object 
$\delta(\widehat{\cal C})$. 
This is not as weird as it might 
look at first sight, because it is actually defined as 
\be
\delta(\widehat{\cal C}) = \int_{-\infty}^\infty 
\frac{d\lambda}{2\pi}\, e^{ i \lambda \widehat{\cal C}} \, , 
\label{delta}
\ee
in a sense to be specified in a moment. 
Whenever this construction makes sense, one expects 
$\widehat{\cal C} \,\delta(\widehat{\cal C}) = 0$, so that, 
once a function $\psi\in {\cal H}$ is specified, the quantity 
\be
\psi_{\rm phys} = \delta(\widehat{\cal C})\,\psi\, ,
\label{psidelta}
\ee
provided it exists, 
should be a solution of the wave equation. 
Of course, this has to be made
more precise. Equation (\ref{delta}) in fact tells us how to 
proceed: By simply {\it first} applying 
$ e^{ i \lambda \widehat{\cal C}} $ to $\psi$ and {\it thereafter}
integrating over $\lambda$, the quantity (\ref{psidelta}) 
becomes an integral that 
has a chance to converge. 
A technically important requirement when doing so is that the 
constraint operator $\widehat{\cal C}$ shall be self-adjoint, 
so that $ e^{ i \lambda \widehat{\cal C}} $ is unitary for 
any real $\lambda$. 
This is in fact the case for 
a wide class of models $({\cal M},g_{\alpha\beta},U)$. 
Moreover, if the spectrum of $\widehat{\cal C}$ is  
purely continuous near zero, 
the definition (\ref{delta}) should intuitively be as reasonable as if 
$\widehat{\cal C}$ was a real variable. 
\medskip 

One would not expect the result $\psi_{\rm phys}$ to be  
a normalizable function (simply because in general there are no
normalizable solutions to the wave equation), but one may expect 
it to be a function on $\cal M$ 
solving (\ref{WDW}) --- at least for reasonable choices of $\psi$. 
\medskip 

Since an element $\psi\in{\cal H}$ does not satisfy (\ref{WDW}) and 
hence does not represent a physical 
state of the system, it is called ``kinematical state'' and
${\cal H}$ is called ``kinematical Hilbert space''. 
Accepting the way (\ref{psidelta}) of writing down the relation 
between kinematical and physical states, we proceed and try 
to compute the inner product between two physical states, 
\be
\psi_{\rm 1,phys} = \delta(\widehat{\cal C})\, \psi_1
\qquad\qquad 
\psi_{\rm 2,phys} = \delta(\widehat{\cal C})\, \psi_2 \, , 
\label{two}
\ee
a quantity about which we know from the outset that it is 
ill-defined:   
Formally, we get 
\be
\langle \psi_{\rm 1,phys}, 
\psi_{\rm 2,phys}\rangle  = 
\langle \psi_1, \delta(\widehat{\cal C})^2 
\psi_2\rangle = \delta(0) \, 
\langle \psi_1 \delta(\widehat{\cal C}) \psi_2\rangle 
\label{ll}
\ee
where I have written $\delta(\widehat{\cal C})^2 = 
\delta (0) \delta(\widehat{\cal C}) $ in order to indicate
where the infinity comes from. 
\medskip 

The key idea with far reaching consequences is now very simple: 
just drop $\delta(0)$! 
Given the two states (\ref{two}), 
define an inner product $\langle\,,\,\rangle_{\rm phys}$ 
as 
\be
\langle \psi_{\rm 1,phys}, 
\psi_{\rm 2,phys} \rangle_{\rm phys}  = 
\langle \psi_1, \delta(\widehat{\cal C})\, \psi_2\rangle \, . 
\label{inner}
\ee
This has a chance to be finite even if the two physical states 
are not normalizable with respect to the 
kinetical inner product $\langle\,,\,\rangle$. 
\medskip 

The equations written down so far provide the motivation 
(and the basis for the exactification) of the RAQ scheme in the case
of a single constraint. 
The set of solutions to (\ref{WDW}) which may be written in the form 
(\ref{psidelta}) and which are normalizable with respect to 
the inner product (\ref{inner}) defines the 
Hilbert space
$({\cal H}_{\rm phys},\langle\,,\,\rangle_{\rm phys})$ 
of physical states. 
\medskip 

There is another way to write down 
the same construction. The quantum constraint (\ref{WDW}) may be 
viewed as the eigenvalue problem of the constraint operator 
$\widehat{\cal C}$ for the eigenvalue zero. If zero was indeed 
an eigenvalue of $\widehat{\cal C}$, then a solution 
$\psi_{\rm phys}$ would just be an element of the kernel of 
$\widehat{\cal C}$. Then one could interpret (\ref{psidelta}) 
as the projection onto this kernel (and 
$\delta(\widehat{\cal C})$ would in fact better be written as
$\delta_{\widehat{\cal C},0}$). However, since 
$\widehat{\cal C}$ has purely continuous spectrum near zero, 
there is no such element in the Hilbert space $\cal H$, and zero 
is only a {\it generalized} eigenvalue. 
One thus needs a generalized sort of ``projection'' 
onto the (generalized) kernel of $\widehat{\cal C}$. 
This throws one {\it out of} the kinematical Hilbert space 
$\cal H$. 
In the approach outlined above the quantity 
$\delta(\widehat{\cal C})$ has played the role 
of such an operation. 
Again, some 
acrobatics with the delta function provides an intuitive
piece of insight. 
Under quite weak conditions there is a set of
generalized eigenvectors $\psi_{j,E}$ of $\widehat{\cal C}$, 
where $E$ is a continuous and
$j$ a discrete index, satisfying 
\begin{eqnarray}  
\widehat{\cal C} \,\psi_{j,E} &=& E\, \psi_{j,E} 
\label{eigen}\\
\langle \psi_{j,E} , \psi_{j',E'}\rangle &=& 
\delta(E-E')\,\delta_{j j'}\, , 
\label{deltaEE}
\end{eqnarray}
and being complete in the sense 
that any $\psi\in{\cal H}$ is given by a linear superposition 
of the form
\be 
\psi = \sum_j \int dE\, c_j(E)\, \psi_{j,E} 
\label{linkomb}
\ee
(I omit measure-theoretical subtleties here). 
The discrete index accounts for the degeneracy of the generalized 
eigenvalues $E$. (One may think of the Fourier transformation 
in $L^2(\real)$ for a well-known example of a basis of generalized 
eigenvectors). 
\medskip

Note that the $\psi_{j,E}$ are not elements of $\cal H$. 
Nevertheless, by virtue of (\ref{eigen}), the functions $\psi_{j,0}$ 
solve the quantum constraint (\ref{WDW}). (This becomes quite clear 
if one likes to compute the $\psi_{j,E}$ in practice: one would 
have to {\it solve} (\ref{deltaEE}) for any $E$ (up to
boundary conditions that essentially exclude exponentially increasing 
functions), the index $j$ just counting 
the linearly independent solutions. 
In this sense, the physical states (i.e. the solutions of the wave 
equation) are already at our disposal. The formal inner product 
(\ref{deltaEE}) between two $E=0$ functions of course 
diverges 
\be 
\langle \psi_{j,0} , \psi_{j',0}\rangle = 
\delta(0)\, \,\delta_{j j'}\, , 
\label{deltaEE0}
\ee
and this can be cured by dropping $\delta(0)$ and {\it defining} 
the physical inner product as 
\be 
\langle \psi_{j,0} , \psi_{j',0}\rangle_{\rm phys} = 
\delta_{j j'}. 
\label{inner2}
\ee
The $\delta(0)$ from (\ref{ll}) and from (\ref{deltaEE0}) 
are in fact the same! Again, in this picture, the
non-normalizability of physical states with respect to the 
kinetical inner product has been overcome by dividing by an 
infinite constant. 
The set $\{\psi_{j,0}\}$ then provides an orthonormal basis 
(in the usual sense) of ${\cal H}_{\rm phys}$. 
In order to make this second approach 
more rigorous one has to make use of the spectral theory of self-adjoint 
linear operators. In fact, this one of Marolf's contributions to the 
RAQ scheme \cite{Marolf2}). 
The relation to the first approach is provided by 
$\psi_{j,E} = \delta(\widehat{\cal C}-E) \chi_{j,E}$ for 
suitably chosen kinematical states $\chi_{j,E}$. For $E=0$, this 
reduces to (\ref{psidelta}). 
\medskip 
 
The framework outlined so far in this Section, 
although at a heuristic level, 
suffices to carry out practical computations. For 
non-trivial cases, the wave equation is of course very difficult
to solve explicitly, and so a closed formula for the physical inner 
product is still out of reach for many (actually most) interesting models. 
But these difficulties are more of practical than of 
conceptual nature. 
\medskip 

An instructive example is that of the flat Klein-Gordon equation 
($g_{\alpha\beta}=\eta_{\alpha\beta}$ flat 
and $U=m^2$ constant, as treated in Section 1). 
Solutions of the wave equation are functions of the 
form (\ref{posneg}), and a more or less straightforward 
application of the formulae written down so far 
(either by computing the action of $\delta(\widehat{\cal C})$ or 
by finding $\psi_{j,E}$) 
reveals that the physical inner product between two 
such functions $\psi_1$, $\psi_2$ is given by
\be
\langle\psi_1,\psi_2\rangle_{\rm phys} = 
4 \pi 
%%% hab ich nachgerechnet 
\int d^3k \Big( a^*_{1}(\vec{k}) a_2(\vec{k}) +  
b^*_{1}(\vec{k}) b_2(\vec{k})  \Big) \, .   
\label{innermink}
\ee
This provides the exact definition of the physical Hilbert 
space ${\cal H}_{\rm phys}\,$: it consists of those functions
of the form (\ref{posneg}) 
which are normalizable with respect to (\ref{innermink}). 
Both $a(\vec{k})$ and $b(\vec{k})$ must be square integrable, so that 
${\cal H}_{\rm phys}$ becomes isomorphic to 
$L^2(\real)\times L^2(\real)$. 
Note that the difference between 
(\ref{KGmink}) and (\ref{innermink}) 
is essentially a sign. The physical inner product 
is thus obtained by reversing the sign of the indefinite 
scalar product (\ref{KG}) in the negative frequency sector. 
Although this may be disturbing, such a reversion of sign is 
perfectly compatible with Lorentz invariance. (In fact, 
this observation was one of the starting points of Higuchi's 
contribution to the RAQ scheme \cite{Higuchi}). 
The positivity of the inner product $\langle\,,\,\rangle_{\rm phys}\,$, 
being desirable from the point of view of
interpretation, provides another major advantage 
over the indefinite scalar product $Q$: 
it leads to a precise definition of the set of states 
to work with, whereas
$Q$ does not (as already remarked in Section 1: 
the set of solutions $\psi$ to the wave equation 
for which $Q(\psi,\psi)$ is finite contains 
asymptotically ill-behaved functions). 
\medskip 

%%% CORR BEGIN 12.8.97 
There is yet a third way to intuitively understand the transition 
from the kinetical to the physical inner product. Classically, 
the constraint function 
(\ref{C}) plays a double role, the first of which --- constraining the
phase space variables and thus leading to the reduced phase space --- 
is carried over to the quantum theory by
imposing the wave equation (\ref{WDW}). However, its second role is
to generate gauge transformations (as any first class constraint does), 
the infinitesimal expression for which 
is given by $\delta f = \epsilon \{ {\cal C}, f\}$ for 
functions $f\equiv f(x,p)$. This induces a flow on the 
reduced phase space, connecting  
points which have to be considered physically equivalent. 
Quantum mechanically, this is accounted
for by inserting a gauge-fixing delta function together with a 
Faddeev-Popov determinant into the integral (\ref{KIN}). 
The kinetical inner product --- being defined without 
this insertion --- contains a redundant piece, namely 
the integration over gauge-orbits. 
It thus differs from the physical inner product by a factor 
which represents the redundant integration. The latter 
is infinite and may in turn be identified with our $\delta(0)$. 
It was stressed by Woodard \cite{Woodard} that this procedure 
--- being standard reasoning in perturbation theory around flat 
space --- 
should be applied in the quantum gravity and quantum cosmology 
context as well\footnote{I became aware of Woodard's beautiful 
paper only after having written up this conference contribution, 
thanks to a hint by Steve Carlip.}. 
%%% CORR END 12.8.97 

A way of interpreting the physical states constructed by the 
RAQ scheme in terms of observations 
has been proposed by Marolf \cite{Marolf3}. 
I will not enter this issue, nor will I say something about how 
classical observables are promoted into 
quantum ones, i.e. self-adjoint linear operators on 
${\cal H}_{\rm phys}$
(see Ref. \cite{Marolf1} for this topic), 
but I will now extract a more general 
structure from the picture developed so far. 
By separating several issues from each other 
it should also become clear how the scheme can be extended 
to more realistic situations. 
\medskip 

\section{Towards the general RAQ scheme --- still hand-waving} 

The RAQ scheme is in fact a list of rules that may sometimes 
be applied more or less straightforwardly 
(as for the simple models considered 
above, given that its ingredient are not too pathological), 
while in more sophisted systems 
crucial ambiguities may appear, so that a lot of choice has to be 
made on the way. In order to outline two major items of this list,
I begin reformulating what we have done above in a slightly more
precise language. 
(I ignore here many problems that arise even {\it before} this point, 
starting from a quite generic classical system, for 
example the very definition of the kinematical Hilbert
space in cases where the analogy of (\ref{KIN}) is not 
obvious; see Refs. \cite{Aetal}\cite{Marolf1}). 
\medskip 

The minisuperspace framework provides 
the kinematical Hilbert space $\cal H$ and the constraint operator 
$\widehat{\cal C}$ as objects to start with. 
In more sophisticated models one will encounter 
a kinematical Hilbert space $\cal H$ as well, together with 
a family of constraint operators (say $\widehat{\cal C}_I$). 
The goal is to ``solve the constraint(s)'' by defining a 
Hilbert space of physical states {\it just in terms of the 
Hilbert space framework}. So let us reconsider the minisuperspace
example, trying to extract a procedure that can be stated 
solely in terms of $({\cal H}, \widehat{\cal C})$ 
and structures contained in this framework. 
\medskip 

First of all we note that the constraint operator 
$\widehat{\cal C}$ 
is an unbounded linear operator, so it can be defined only on a 
dense subspace ${\cal D}(\widehat{\cal C})$ of $\cal H$, 
\be
\widehat{\cal C} : {\cal D}(\widehat{\cal C}) 
\rightarrow {\cal H} \, . 
\label{densesubspace}
\ee
Nevertheless, if it is self-adjoined, 
the operators $e^{i\lambda \widehat{\cal C}}$ are unitary and 
hence well-defined on the whole of $\cal H$. 
\medskip 

Almost everything said in the preceding Section  
cries for an interpretation of the physical states in terms 
of distributions. 
The object $\delta(\widehat{\cal C})\,\psi$ --- which I have 
argued to be a physical state, i.e. a function 
$\psi_{\rm phys}$ on $\cal M$ 
satisfying the quantum constraint (\ref{WDW}) --- is not 
normalizable with respect to $\langle\,,\,\rangle$, hence it is 
not an element of the kinematical Hilbert space ${\cal H}$. 
However, the quantity 
$\langle \delta(\widehat{\cal C})\,\psi, f\rangle$, 
when witten as an integral, using (\ref{delta}), 
has a chance
to be finite for at least a large set of elements 
$f\in{\cal H}$ (just as the integral
$\int dx\, f(x)$ has a chance to be finite for 
certain elements 
$f\in L^2(\real)$). In general, the introduction of 
distributions on a Hilbert space
may be motivated by the attempt to 
give the scalar product between a non-normalizable object 
(distribution) and a 
normalizable object (test function) a precise meaning. 
In our case, given $\psi$, we encounter the linear assignment 
\be
\eta(\psi):
f\,\mapsto\, \langle \delta (\widehat{\cal C})\,\psi,f\rangle = 
\int_\infty^\infty \frac{d\lambda}{2\pi}\, 
\langle \psi, e^{- i \lambda \widehat{\cal C}} f\rangle\, . 
\label{assign}
\ee
Here, $\psi$ stands for a kinematical state and $\eta(\psi)$ 
is a linear map from (a subspace of) ${\cal H}$ into 
the complex numbers. {\it This map} encodes all 
the information I have 
naively written down as $\delta(\widehat{\cal C})\,\psi$, so that the 
idea is to {\it consider it as a physical state}. If this can be
made rigorous, $\eta$ would be an anti-linear map sending 
(certain) kinematical states to physical states. 
\medskip 

This means that we re-interpret the {\it functions} $\psi_{\rm phys}$ 
satisfying (\ref{WDW})  
as {\it maps} $\eta(\psi)$ of the type (\ref{assign}).  
In this way physical states become distributions. 
These distributions may of course still be {\it represented as 
functions} $\psi_{\rm phys}$ 
(the relation between the function $\psi_{\rm phys}$ 
and the distribution $\eta(\psi)$ 
being equation (\ref{assign}), when written in the
form $\eta(\psi):f\mapsto \langle\psi_{\rm phys},f\rangle$, 
which is just an integral over $\cal M$ of the form (\ref{KIN})), 
so that the arguments of the preceding Section 
are not invalidated. However, the new point of view is 
generalizable since it develops the notion of non-normalizable 
objects {\it from the Hilbert space itself} and does not need to 
refer to specific properties of the model. 
\medskip 

What spaces are $\psi$ and $f$ allowed to come from? 
After all, the expression (\ref{assign}) shall be finite. 
If a large subspace of $\cal H$ is allowed to inhabit the possible
$\psi$'s, there will be only few $f$'s for which 
(\ref{assign}) exists. Conversely, if $\psi$ is 
restricted to come from a small subspace, then we will find 
many $f$'s for which (\ref{assign}) exists. 
Here things become ambigous, although any reasonable choice 
will yield the same final result for our simple 
minisuperspace models. 
However, instead of searching a 
preferred choice of these spaces just for the models considered
so far, we anticipate more sophisticated situations and 
try to extract the crucial contents of the present situation. 
Let us denote by $\Phi$ the subspace of $\cal H$ in which $\psi$ 
has to lie, and suppose it to be dense in $\cal H$.  
\medskip 

The {\it first} major choice in the RAQ program is thus to 
fix a dense subspace $\Phi$ of $\cal H$. 
Moreover, the space $\Phi$ shall be a subspace of 
${\cal D}(\widehat{\cal C})$ and it shall be invariant under 
the action of the 
constraint operator $\widehat{\cal C}$. 
Let $\Phi'$ be the (topological) dual of $\Phi$, i.e. the 
space of all continuous linear map from $\Phi$ into 
the complex numbers. 
This is the space of distributions which provides a home 
for the generalized 
eigenvectors of $\widehat{\cal C}$ 
and inside which we will identify the physical states. 
In order to have a non-trivial notion of continuity, we must 
assume a topology on $\Phi$ which is finer than the 
Hilbert-space topology. It is usually assumed to be a 
nuclear topology. For simple models like (\ref{WDW}), one does
not really need to postulate so much freedom, but in complicated 
models this might be necessary (although the last word 
has certainly not yet been spoken on this issue). 
Anyway, given such a suspace, one may 
imbed $\cal H$ in $\Phi'$ (because taking the inner product with
a fixed element of $\cal H$ defines an element of $\Phi'$), 
so that one gets the chain of inclusions 
$\Phi \subset {\cal H} \subset \Phi'$ (which is called a 
Gelfand triple). 
\medskip 

The elements of $\Phi$ will play the role of test functions. 
Possible candidates for $\Phi$ to start with in minisuperspace 
models are thus of the Schwarz or $C_0^\infty$ type, whereas 
in complicated situations, e.g. in all attempts to treat 
full gravity, the choice of $\Phi$ can be 
a difficult task. 
\medskip 

The {\it second} major choice to be made in the RAQ program 
is to fix an anti-linear map (called ``rigging map'') 
\be
\eta: \Phi\rightarrow \Phi' 
\label{eta}
\ee 
such that $\eta(\psi)$ ``satisfies the constraint'' 
for any $\psi\in\Phi$.   
What does this mean? In the simple models considered above,  
the constraint is solved because the quantity
$\delta(\widehat{\cal C})$ is floating around. 
Equation (\ref{assign}) is just the definition of $\eta$, once 
the question of domains has been fixed. 
It may be interpreted by stating that the constraint 
operator $\widehat{\cal C}$ generates a group 
(of unitary operators) over which an average 
is performed. (Hence the 
alternative name ``group average'' for RAQ. 
%%% CORR BEGIN 12.8.97
There should, by the way, also be relations to the 
Faddeev-Popov way to define the physical inner product mentioned
in Section 2). 
%%% CORR END 12.8.97
This provides a hint how to proceed in case of more than just 
one constraint: Consider the group (of unitary operators) 
generated by the constraints and carry out the average. 
The average over a compact group is provided by the 
(unique) Haar measure, in which case (\ref{assign}) 
finds a natural generalization. However, in case a non-compact 
group appears, there is no general prescription, and the 
problem of averaging must be tackled for any particular 
model. (For example, since the diffeomorphism group 
on a manifold is not compact, the diffeomorphism constraint in 
loop quantum gravity is of this type 
\cite{Aetal}\cite{MarolfMouraoThiemann}). 
\medskip 

In order not to refer to an average procedure that might not 
exist in certain models, I rather state the {\it goal} which 
has to be achieved and for which the group average 
is just a tool: solving the constraint(s). 
Since we are now considering 
a physical state to be an element 
$\eta(\psi)\in \Phi'$ (stemming from an 
element $\psi\in\Phi$), the statement that it solves 
the constraint is reformulated as 
\be
(\eta(\psi))\,\widehat{\cal C} f = 0 
\qquad\qquad \forall\, f \in \Phi\,  . 
\label{satconst}
\ee
Note that, since $\Phi$ is left invariant by 
$\widehat{\cal C}$, this equation makes perfect sense. 
Thus, expressed in Dirac's notation, one does not actually solve
the equation 
$\widehat{\cal C}\,| \psi\rangle=0$ but rather 
the equation 
$\langle \psi |\,\widehat{\cal C} =0$ in a distributional 
sense. (This reflects, by the way, the anti-linear nature of $\eta$). 
In the minisuperspace models this condition is identical to 
$\psi_{\rm phys}$ satisfying the wave equation
(just partially integrate 
$\langle\psi_{\rm phys},\widehat{\cal C} f\rangle=0\,\forall 
f\in\Phi$), so that 
the hand-waving framework of the preceding Section is recovered. 
Also, the generalization of (\ref{satconst}) to the case of 
several constraints is obvious, once their action is known. 
(All constraints shall act on $\Phi$ and leave $\Phi$ 
invariant). 
In general, whenever a group average exists, the resulting 
map $\eta$ satisfies this property. If a group average 
is not obviously defined, one might 
try to achieve (\ref{satconst}) for all $\psi\in \Phi$ 
by any other method at hand. 
In order to mention an example: for the diffeomorphism constraint 
in loop quantum gravity
(which is actually an uncountable family of constraints), 
it is an easier task to realize 
an appropriate version of (\ref{satconst}) than 
to try to generalize the notion of group average. 
\medskip 

There are some further requirements for $\eta$, among which the 
most important ones are 
$(\eta(\psi_1))\psi_2 = ((\eta(\psi_2))\psi_1)^*$ and
$(\eta(\psi))\psi\geq 0$. All this may easily be satisfied for 
the models of the type (\ref{WDW}) but prevents potential 
difficulties for more complicated scenarios. 
\medskip 

Once having succeeded to fix $\Phi$ and $\eta$, 
only some simple definitions remain to be done. 
The physical Hilbert space is defined as the completion of 
$\eta(\Phi)$ with respect to the inner product 
\be
\langle \eta(\psi_1), \eta(\psi_2) \rangle_{\rm phys} = 
(\eta(\psi_2)) \,\psi_1 
\label{inner3}
\ee
for $\psi_1,\psi_2\in \Phi$. 
Note that $\psi_1$ and $\psi_2$ have interchanged position 
at the right hand side. 
This stems from the anti-linearity of $\eta$, together with the 
convention that 
$\langle\,,\,\rangle_{\rm phys}\,$ shall be anti-linear in the 
first factor. 
The fact that there is just one $\eta$ at the right hand side 
is the exactified version of dropping an infinite 
factor $\delta(0)$ as was done in 
Section 2. 
In this way the space of 
physical states emerges on a rather abstract level 
as a certain set of distributions solving the constraint(s) 
in a certain sense. 
In the minisuperspace models, as already anticipated, 
the elements of $\eta(\Phi)$ 
may be represented as functions $\psi_{\rm phys}$ 
on $\cal M$ satisfying the wave equation
(at least for a reasonable choice of $\Phi$), whereas 
passing over to the completion adds 
objects which satisfy (\ref{WDW}) only in a Hilbert space
sense (thus implying a weaker notion of differentiability). 
In more complicated models, such an intuitive procedure 
might not be possible, in which case one will be obliged 
to work entirely at the abstract level 
in order to arrive at ${\cal H}_{\rm phys}$. 
\medskip 

This is not the end of the story. I have ignored the problem of 
observables and many technicalities. 
In a realistic system like full gravity, any step in the list 
(including those I did not even mention) provides its own problems. 
Despite the individuality and complexity of these problems, 
there are some common underlying ideas, encoded in the RAQ scheme, 
some of which I (hopefully) have made clear.

\end{document}